\documentclass[letterpaper]{aa}                                         % DOCUMENT CLASS AND ABSTRACT TYPE

 \usepackage{epsfig,graphicx,natbib,amsmath}    %PACKAGES INCLUDED

\bibpunct{(}{)}{;}{a}{}{,}                                                               %  fix to natbib
\begin{document}

\title{Unusual Stokes V profiles during flaring activity of a delta sunspot}

\author{C.E.~Fischer\inst{1} \and C.U.~Keller\inst{2} \and F.~Snik \inst{2} 
 \and  L.~Fletcher \inst{3}  \and H.~Socas-Navarro \inst{4}}
 
\institute{ ESA/ESTEC RSSD, Keplerlaan 1, 2200 AG Noordwijk, The Netherlands \and  Sterrewacht Leiden, Universiteit Leiden, Niels Bohrweg 2, 2333 CA Leiden, The Netherlands\and School of Physics and Astronomy, SUPA, University of Glasgow, Glasgow G12 8QQ,UK  \and Instituto de Astrof\'{i}sica de Canarias, Avda V\'{i}a L\'{a}ctea S/N, La Laguna 38200, Tenerife, Spain}

\date{Received / Accepted }

\abstract {}{We analyze a set of full Stokes profile observations of the flaring active 
region NOAA 10808. The region was recorded with the Vector-Spectromagnetograph (\emph{VSM}) of the \emph{SOLIS} facility. The active region produced several successive X-class flares 
between 19:00 UT and 24:00 UT on September 13, 2005 and we aim to quantify transient and permanent changes in the magnetic field and velocity field  during one of the flares, which has been fully captured.
} {The Stokes profiles were inverted using the height-dependent inversion code LILIA to analyze magnetic field vector changes at the flaring site. We report multilobed asymmetric Stokes $V$ profiles found in the $\delta$-sunspot umbra. We fit the asymmetric Stokes $V$ profiles assuming an atmosphere consisting of two components (SIR Inversions) to interpret the profile shape.  The results are put in context with \emph{MDI} magnetograms and reconstructed \emph{RHESSI} X-ray images.}
{We obtain the magnetic field vector  and find signs of restructuring of the photospheric magnetic field during the flare close to the polarity inversion line (PIL) at the flaring site. At two locations in the umbra we encounter strong fields ($\sim$\,3\,\rm{kG}), as inferred from the Stokes $I$  profiles which, however, exhibit a low polarization signal. During the flare we observe in addition asymmetric Stokes $V$  profiles at one of these sites. The asymmetric Stokes $V$ profiles appear co-spatial and co-temporal with a strong apparent polarity reversal observed in \emph{MDI}-magnetograms and a chromospheric hard X-ray source.
The two-component atmosphere fits of the asymmetric Stokes profiles result in line-of-sight velocity differences in the range of  $\sim$\,12\,$\ensuremath\mathrm{km}/{\mathrm{s}}$ to 14\,$\ensuremath\mathrm{km}/{\mathrm{s}}$ between the two components in the photosphere. Another possibility is that local atmospheric heating is causing the observed asymmetric Stokes $V$ profile shape. In either case our analysis shows that a very localized patch of $\sim$\,5\,$\arcsec$ in the photospheric umbra, co-spatial with a flare footpoint, exhibits a sub-resolution fine structure.} {}

\titlerunning{Unusual Stokes profiles}
\authorrunning{C.E.Fischer et al.}
\keywords{Sun: sunspots, Sun: photosphere, Sun: flares, Sun: surface magnetism}
\maketitle

%============================================================================================================================================
   %INTRODUCTION

%============================================================================================================================================

\section{Introduction}     \label{sec:introduction}

It is generally believed that flares are the result of the release of stored magnetic energy during a reconnection process accompanied by a change of the magnetic topology~\citep[see e.g.][]{1964NASSP..50..451C, 1966ApJ...143....3S, 1976SoPh...50...85K}. Photospheric magnetic shear and disturbance of already emerged magnetic loops by new flux emergence are suspected to play an important role in the energy loading process, see e.g.,~\cite{2002ApJ...577..501K},~\cite{2011ApJ...726...12E} or the recent review by~\cite{lrsp-2011-6}. 

The back reaction of a large energy release on the lower atmosphere has been observed as photospheric `sunquake' \citep{{1998Natur.393..317K}, {2005ApJ...630.1168D}}, non-reversing changes in the line-of-sight magnetic field \citep{{1999ApJ...525L..61C},{2005ApJ...635..647S}}, and heating of the chromospheric temperature minimum and upper photosphere \citep{1974SoPh...38..499M,1990ApJ...365..391M,2004ApJ...607L.131X}. It is clear that the effects of the enormous, yet localized, release of stored magnetic energy in the corona has dramatic impacts on the atmosphere beneath, but the manner of the observed response of the deep atmosphere remains a puzzle. The primary agents for carrying energy in flares are thought to be fast electrons, but these cannot penetrate deep enough in the atmosphere, in sufficient number, to directly generate the heating seen \citep{1986A&A...156...73A}. Fast ions have a better penetration range, but though these are almost certain to exist in large numbers in flares, they are difficult to observe. In any case the better association of both sunquakes and chromospheric/photospheric heating seems to be with electrons \citep{2007ApJ...656.1187F,2007ApJ...670L..65K}. Heating of the mid- or upper chromosphere resulting in the emission of UV radiation that then causes photospheric `backwarming' \citep{1989SoPh..124..303M} is also likely.  Energy may also be carried by magnetic disturbances, which have the potential to penetrate deep into the atmosphere \citep{2008ApJ...675.1645F}, but their dissipation and heating is as yet unknown. 

One of the major obstacles in the study of the connection of photospheric magnetic footpoints and flare events, is the lack of observations of the full magnetic vector in the lower photosphere during a flare.  So far only a few full Stokes profile measurements were recorded covering a flare e.g.,~\cite{2009A&A...505..771H} (\emph{Swedish Solar Telescope}),~\cite{2007PASJ...59S.779K} (\emph{HINODE}) and recently \cite{2012ApJ...745L..17W} (\emph{Solar Dynamics Observatory}). Covering a larger flare site as is often required in M- and X-class flares one suffers some penalty in either spatial, spectral or temporal resolution.
Only with sufficient spectral sampling in the Stokes profiles can one infer sub-resolution structures or draw conclusions on strong gradients in the atmospheric parameters from the shape of the profiles \citep{1998A&A...336L..65F,2001ApJ...563.1031S,2006A&A...460..925M,2011A&A...530A..14V}. For example, \cite{2012arXiv1201.6312K}, revealed a complex multicomponent atmosphere around flare footpoints by analyzing high resolution \emph{IBIS} (\emph{Interferometric BIdimensional Spectrometer}) data.

The rare full Stokes data set presented here gives us the opportunity to investigate the magnetic field vector changes associated with a flare, but due to the high spectral resolution, also allow a more detailed analysis of the Stokes profile shapes. This type of detailed analysis offers the possibility of understanding not only the flare magnetic field evolution at the photospheric level, but also heating and flows in the deep atmosphere. 

%==========================FIGURE 1 : MAGNETOGRAM / CONTINUUM
 \begin{figure}
 \resizebox{\hsize}{!}{\includegraphics{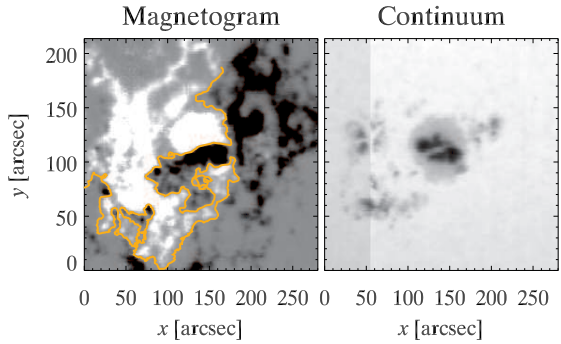}}
  \caption{ Active region NOAA 10808 on September 13, 2005.
    {\em Left:\/} Magnetogram obtained from \emph{SOLIS-VSM} Stokes $V$ profile amplitudes 
in the Fe\,{\sc I} 630.25\,\rm{nm} line. The white areas denote positive magnetic flux, whereas 
the black areas are regions exhibiting negative magnetic flux. The orange line shows the intricate polarity 
inversion line (PIL). 
{\em Right:\/} Co-spatial intensity image constructed from the Stokes $I$ continuum intensities. 
}
 \label{fig:magcont}
\end{figure}
%=================================================
%INTRODUCTION DATA SET

 %=============================FIGURE 2: TIMING GOES / FLARE CLASSIFICATION
\begin{figure}
 \resizebox{\hsize}{!} {\includegraphics{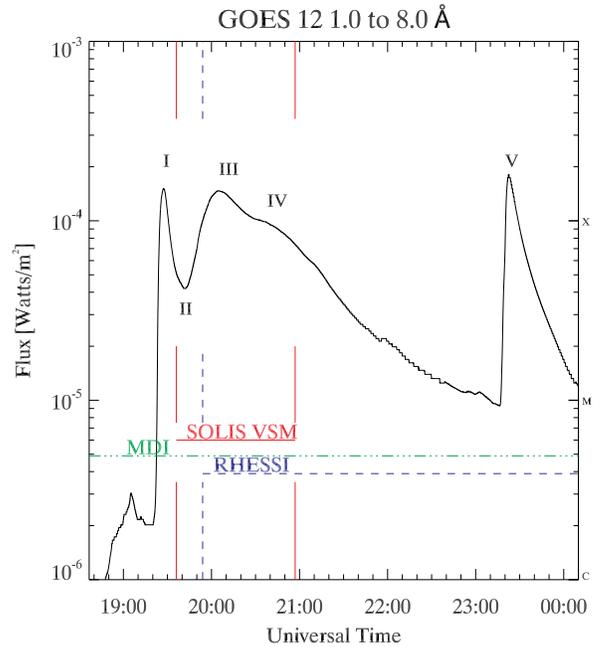}  }
 \caption[]{GOES X-ray flux recorded on September 13, 2005. The flare numbers, adapted from 
\cite{2009ApJ...703..757L}, are indicated with capital roman numbers. The second flare is 
not discernible in the soft X-ray flux. The \emph{SOLIS-VSM} observation time from 19:36 UT to 20:57 UT is represented by the red solid line.
 \emph{MDI}, recording magnetograms in Ni\,{\sc I} 676.77\,\rm{nm} and the \emph{RHESSI} observing times are indicated with green dashed-dotted and blue dashed lines respectively. }
\label{fig:timing} 
\end{figure}

%\subsection {NOAA 10808 on September 13, 2005}
 %======================================================================
%FLARE OBSERVATIONS
The observed active region, NOAA 10808, showed a complex 
structure and is a so-called $\delta$-spot, as its two opposite polarity umbrae share a common penumbra.
Figure~\ref{fig:magcont} shows the magnetogram and the continuum intensity of the 
active region. The intricate polarity inversion line (PIL) indicates the high
complexity of the magnetic configuration typical for a $\delta$-sunspot. NOAA 10808 
displayed a high flare activity, producing over 10 X-class flares and many M-class 
and C-class flares during its presence on the observable solar disk. 
On September 13, 2005 the active region was located almost at the center of the
solar disk and the \emph{Geostationary Operational Environmental Satellites (GOES)} recorded 
a succession of soft X-ray (SXR) flux peaks within 4.5 hours. Figure~\ref{fig:timing} shows 
the SXR flux registered by \emph{GOES} and indicates the \emph{SOLIS} (\emph{Synoptic Optical Long-term Investigations of the Sun}) \emph{VSM} 
observation time, as well as the observation time of \emph{RHESSI}(\emph{Reuven Ramaty High Energy 
Solar Spectroscopic Imager}). \emph{MDI} (\emph{Michelson Doppler Imager}) magnetograms are available for the entire time section.
The flare numbering is adopted from \cite{2009ApJ...703..757L} who analyzed the flares\,{\sc I - IV} using a wide range of wavelength bands. They were 
particularly interested in the evolution of successive flares and concluded that the events were interconnected. 
They identify the delta spot umbra as the location of flare\,{\sc III}. They support their view by analyzing \emph{RHESSI} hard X-ray (HXR) reconstructed images.  
Flare \,{\sc II} cannot be identified in the \emph{GOES} SXR flux. However, \cite{2007ApJ...671..973W} observed a
clear flare signature of two \mbox{H\hspace{0.1ex}$\alpha$} kernels moving apart associated with a filament eruption. 
There have been in addition several studies of the individual flares by e.g.,~\cite{2007ApJ...668..533N},~\cite{2007A&A...472..967C} and~\cite{2009ApJ...693L..27C} as well as comparisons with simulations of the large-scale emergence of a twisted flux rope~\citep{2010A&A...514A..56A}.

%============================================================================================================================================
   %OBSERVATIONS

%============================================================================================================================================

\section{Observations}    \label{sec:dataana}

%DATA REDUCTION
\subsection{Data reduction}
The Stokes profiles were recorded by the ground-based \emph{SOLIS} 
Vector-Spectromagnetograph \citep{2003SPD....34.2023K} on September 13, 2005. 
The \emph{SOLIS-VSM} has a polarization sensitivity of a few 10$^{-4}$ per pixel in less than one second and is capable of obtaining the full Stokes 
profiles for a limb-to-limb slice of $\sim$\,337\,$\arcsec$ in width at a cadence of 5 minutes. The scans are performed by slit scanning with the slit positioned in the geocentric West-East direction and a stepwidth of 1.125\,$\arcsec$. 
The pixel size is 1.125\,$\arcsec$ and the spectral window encompasses the two Fe\,{\sc I} lines at 630.15~\rm{nm} and 630.25~\rm{nm} with a
resolution of 0.0027~\rm{nm} per pixel.
The dark currents are subtracted from the raw data and a flat field correction is employed. The polarization
demodulation matrix is applied to obtain Stokes $I$, $Q$, $U$ and $V$ and fringes are removed. 
All profiles are normalized to the continuum of the quiet Sun intensity, which is obtained
by averaging the continuum values of 10 quiet Sun pixel positions around disk center.
We correlate the magnetograms thus obtained to counteract the spatial drift present between successive active region scans.
Unfortunately the measurement of the polarized spectrum was in some parts severely affected by the deterioration of the polarization modulator. The ferro-electric liquid crystal (FLC) polarization modulator package is located behind the entrance slit. The deteriorating effect of the FLC is the not uncommon formation of limited areas that do not show the expected half-wave retardation and/or fast axis orientation change. While the polarization calibration corrects for these effects, it cannot compensate for the loss in polarimetric efficiency. Therefore, the retrieved Stokes vector in the affected areas is still correct, but the noise is amplified dramatically. As the FLC is close to the entrance slit, the affected area is also limited in the spatial direction along the slit. The profiles in these regions (eastern part of the active region) remained noisy throughout the time sequence. Fortunately large parts of the umbra of the $\delta$-spot are unaffected.

%INVERSIONS
\subsection{Inversion of entire region}
%change text
We use the Levenberg-Marquardt algorithm based code LILIA \citep{2001ASPC..236..487S} to perform the inversions and obtain the global magnetic field vector map. LILIA  assumes local thermodynamic equilibrium and uses a height dependent model with temperature, microturbulence, macroturbulence, magnetic field strength, line-of-sight velocity, magnetic inclination and magnetic azimuth as the atmospheric fit parameters. A (non-polarized) stray-light profile can be provided to the code. This profile can represent the contribution of a non-magnetic atmospheric component in the resolution element or/and instrumental stray-light contaminating the observation. The code then determines a filling factor of this stray-light profile which will be from now on referred to as the stray-light factor.
We used the averaged Stokes $I$ profile of 10 quiet Sun pixels located close to the active region as the stray-light profile. Semi-empirical models obtained by~\cite{2007ApJS..169..439S} are used as initial model atmospheres for the umbra, penumbra and plage profiles correspondingly. 
We apply the non-potential magnetic field calculation method of \cite{2005ApJ...629L..69G} to resolve the 180\,$\degr$ disambiguation in the azimuth angle of the magnetic field vector. In this code the vertical electric current J$_{z}$ is iteratively minimized. We use J$_{z}$ retrieved from the disambiguation of the first time scan as an initial proxy for J$_{z}$ in the following time step and carry on in this way, always using the previous values for the initialization, to guarantee consistency between the time steps.

 %FITTING
 \subsection{Asymmetric profile fits}
 \label{asymfitSIR_GAU}
 As LILIA uses a height-dependent model atmosphere it goes beyond Milne-Eddington Inversions~\citep{1987ApJ...322..473S} and is therefore theoretically able to fit asymmetries in the Stokes $V$ profiles. However, it uses only one magnetic atmospheric component and therefore the complicated profiles caused by superposition of shifted profiles and opposite polarity components in a single spatial resolution element cannot be fitted. The SIR code~\citep{lbelrub2003,1992ApJ...398..375R} is capable of using two magnetic components as well as an additional stray-light component and we employ it to invert the multilobed asymmetric profiles. We can not, however, achieve a satisfactory fit for all of the asymmetric profiles. For these remaining cases we therefore do not perform an inversion and use instead a simple approach by constructing \emph{anti}-symmetric Stokes $V$ profiles consisting of two gaussians with opposite signs. We allow the synthetic \emph{anti}-symmetric Stokes $V$  profiles to be shifted with respect to each other 
 and to carry an optional sign reversal and vary their line width and amplitude. 
 We use these parameters in a fitting routine employing non-linear least squares minimization to determine the best fit to the observed asymmetric Stokes $V$ profiles.

  %===================================================FIGURE 3 : RHESSI LIGHTCURVE
\begin{figure}
\resizebox{\hsize}{!} {\includegraphics {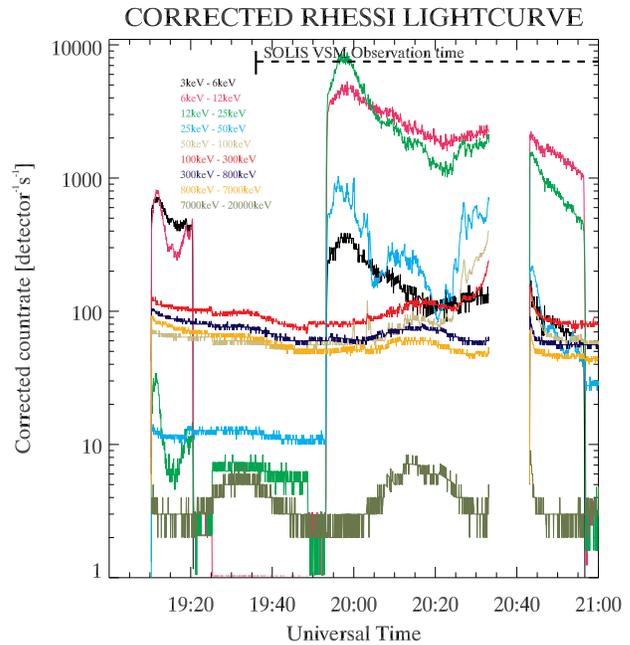}  }
  \caption[]{
\emph{RHESSI} light curves on September 13, 2005. The light curves were corrected for  attenuator and decimation state changes. The rate is shown for various energy
  bands and is an average over 6 detectors. The \emph{SOLIS} observation time is indicated with a dashed horizontal line.  }
  \label{fig:rh_lc} 
\end{figure}
%===========================================================================

%=====================================FIGURE 4: TRACE WL, MAGNETO (PI - P IV),INC and B
\begin{figure*}
\centering
\includegraphics[width=17cm]{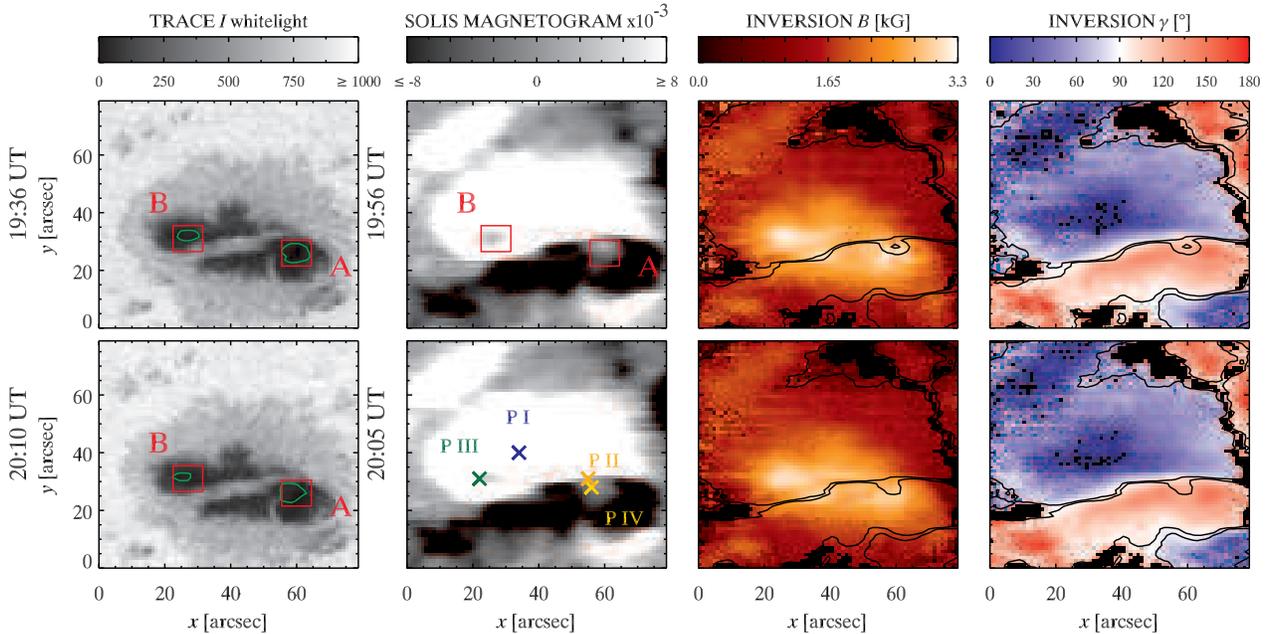} 
  \caption[]{
    Umbral region of NOAA 10808, before and after flare\,{\sc III}. {\em Top\/} The \emph{TRACE} whitelight  intensity image on the left side in the upper row was recorded at 19:36 and rotated and scaled to the \emph{SOLIS-VSM} data. The green contours are locations where the continuum intensity is below $\sim$\,20\,$\%$ of the quiet Sun continuum. These regions marked as A and B are also locations of low polarization signals. The second panel in the upper row is the \emph{SOLIS} magnetogram which was  obtained from the Stokes $V$ profiles (see main text for details) recorded around 19:56 UT. The last two panels, magnetic field strength in kG and inclination angle $\gamma$, are the result of LILIA inversions of the \emph{SOLIS-VSM} full Stokes profiles recorded around 19:56 UT at optical depth log$(\tau_{\rm 500nm})={\rm -1}$. The inclination angle is shown in degrees, with a 
   magnetic field vector along the line of sight pointing out of the solar surface having an inclination angle of 0\degr\,. The black lines are contours obtained by choosing cutoff levels of 0.002 and $-$0.002 for the magnetogram.  The black areas in the inversion results are non-inverted pixels. The noise produced by the deteriorating polarization modulator propagates into the inversion results. {\em Bottom\/} The first panel shows the \emph{TRACE} whitelight image at 20:10 UT. The last three panels in the lower row 
are the magnetogram and the inversions results obtained from \emph{SOLIS} data recorded around 20:05 UT. The Stokes profiles found at locations P\,{\sc I} to P\,{\sc IV} are discussed in section~\ref{subs:tempdev}.} 
\label{fig:invpanels} 
\end{figure*}
%===========================================================================

%======================================FIGURE 5: CHANGE IN INC AND AZIMUTH ANGLE
\begin{figure*}
\sidecaption
  \includegraphics[width=12cm]{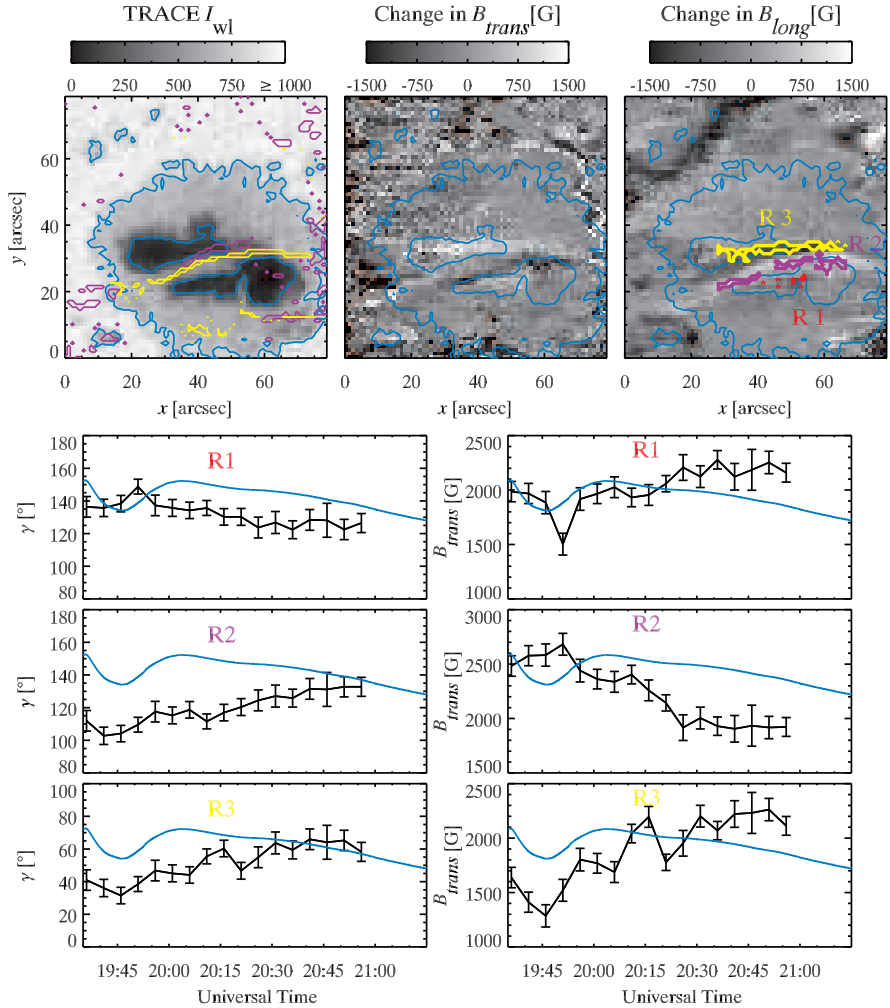} 
 \caption[]{ {\em Top row\/} \emph{TRACE} whitelight image. Purple contours show areas with changes in the magnetic field azimuth larger than 5$\degr$ and yellow contours are areas in which  the longitudinal magnetic field switches from a positive sign to a negative sign or the other way around. All values are retrieved through the inversion and are selected at optical depth log$(\tau_{\rm 500nm})={\rm -1}$. The second and third panels show the change in $B_{\rm trans}$ and $B_{\rm long}$. White areas are regions of increase in $B_{\rm trans}$/ $B_{\rm long}$ and areas of decrease are black. R1 (red contour), R2 (purple contour) and R3 (yellow contour) in the last panel are locations of simultaneous change in $B_{\rm trans}$ and $B_{\rm long}$ greater than 400\,G in both parameters. {\em Last three rows\/} Average inclination angle $\gamma$ and the average $B_{\rm trans}$ in the regions R1, R2 and R3. As in  Fig.~\ref{fig:invpanels} the inclination angle runs from 0\,$\degr$ to 180\,$\degr$ with ${\rm {\gamma} =  90\,{\degr} }$ representing a transversal magnetic field and an angle of  ${\rm \gamma > 90\,\degr }$ constituting a negative longitudinal magnetic field (vector pointing toward the Sun). The error bars are the standard deviation in  $\gamma$ and $B_{\rm trans}$ over a region in the umbra outside the flare site. The soft X-ray \emph {GOES} curve is overlaid in blue.} 
  \label{fig:azy} 
\end{figure*}
%===========================================================================

%======================================FIGURE 6: 4 DIFFERENT STOKES PROFILES
\begin{figure*}
\sidecaption
%\centering
  \includegraphics[width=12cm]{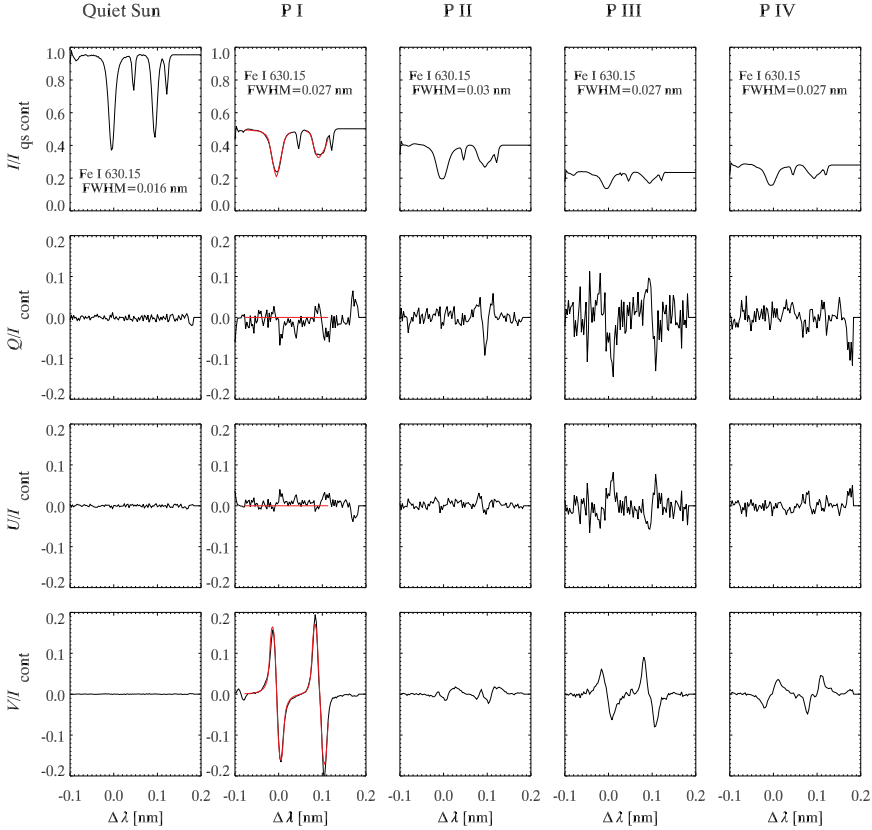}  
    \caption[]{Quiet Sun profile and Stokes profiles retrieved from the positions P\,{\sc I} to P\,{\sc IV} marked in 
Fig.~\ref{fig:invpanels}. The columns show Stokes $I$ with the two Fe\,{\sc I} 630\,nm lines and two telluric oxygen lines normalized to an averaged quiet Sun Stokes $I$ continuum to demonstrate the continuum intensity differences and Stokes $Q$, $U$ and $V$ profiles normalized to the local Stokes $I$ continuum to show the differences in scales between the polarization signals. 
The wavelength scale is given in $\Delta$\,$\lambda$ from the Fe\,{\sc I} 630.15\,\rm{nm} rest wavelength position. The Full-Width-Half-Maximum is given for the Fe\,{\sc I} 630.15\,\rm{nm} line. The red curve over-plotted on the profiles located at P\,I, are synthesized profiles fitted to the observations by the inversion code. The magnetic and velocity field values at log$(\tau_{\rm 500nm}={\rm -1}$) corresponding to this fit are $B$=2.1\,{\rm kG} with an inclination angle of $\gamma$=25$\,\degr$ and a velocity of $v_{LOS}$=$-$1.5\,$\ensuremath{\mathrm{km}/\mathrm{s}}$.}
\label{fig:allprofs}
\end{figure*}
%===========================================================================

%======================================FIGURE 7: TIME DEVELOPMENT ALL PROFS.
\begin{figure*}[!htbp]
%\sidecaption
\centering
  \includegraphics[width=17cm]{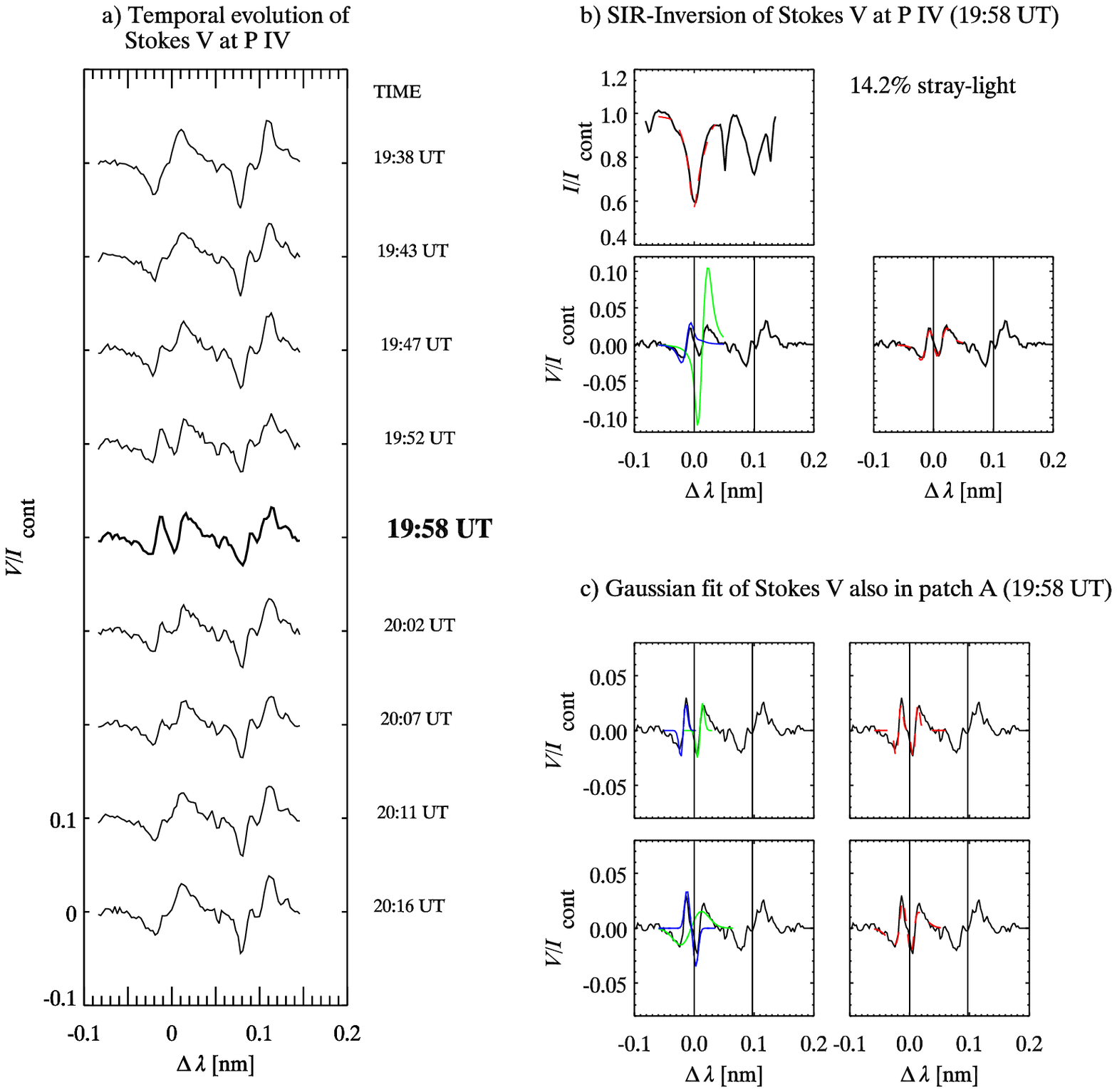} 
 \caption[]{(a) Temporal evolution of Stokes $V$ profile of pixel P\,{\sc IV} located in patch A.  (b) 2-magnetic component SIR Inversion of Stokes profiles at P\,{\sc IV}  at 19:58 UT. The wavelength scale is given in $\Delta$\,$\lambda$ from the {Fe\,{\sc I} 630.15\,\rm{nm}} line rest wavelength position. The upper panel shows the observed Stokes $I$ profile and the Stokes $I$ profile retrieved from the inversion (red dashed line). In the lower panels on the right side we plot the observed Stokes $V$ profile (black solid line) and the Stokes $V$ profile from the best-fit inversion (red dashed line). The rest wavelengths for both {Fe\,{\sc I} 630.15\,\rm{nm}} and {Fe\,{\sc I} 630.25\,\rm{nm}} are indicated by vertical lines. On the left side the Stokes $V$ profiles synthesized from the two individual magnetic atmospheres are shown in green and blue (without taking their respective filling factors into account). The atmospheric parameters from this inversion can be found in Table~\ref{sirtab}. For this fit the inversion resulted in a stray-light percentage of  $\sim$\,14.2\,$\%$. The line-of sight velocity difference between the two components is $\sim$\,12\,$\ensuremath{\mathrm{km}/\mathrm{s}}$. (c) Simple gaussian fit to the observed Stokes $V$ profile of a pixel close to P\,{\sc IV} and also located in patch A. The slit scan containing this pixel was taken approximately 1\,s earlier than the profile at P\,{\sc IV}. The rest wavelengths for both {Fe\,{\sc I} 630.15\,\rm{nm}} and {Fe\,{\sc I} 630.25\,\rm{nm}} are indicated by vertical lines. On the left side we show two \emph{anti}-symmetric Stokes $V$ profiles which were constructed with gaussians and shifted with respect to each other. The velocities in the two components differ by $\sim$\,13.3\,$\ensuremath{\mathrm{km}/\mathrm{s}}$. On the right the combination of the two synthetic \emph{anti}-symmetric Stokes $V$ profiles is shown as a red dashed line. {\em Lower row:\/} On the left side we show a synthetic negative polarity Stokes $V$ profile (green line) and a reversed polarity Stoke $V$ profile (blue). On the right side we show again the combination of the two synthetic \emph{anti}-symmetric Stokes $V$ profiles as a red dashed line. The velocities in the two components differ by $\sim$\,0.48\,$\ensuremath{\mathrm{km}/\mathrm{s}}$ in this case.}   
\label{fig:temp1} 
\end{figure*}
%===========================================================================

%CONTEXT DATA
\subsection{Context data}
\label{cdata}
We retrieve data from \emph{RHESSI} and from \emph{MDI} onboard \emph{SOHO} 
to complement our analysis. 
The \emph{MDI} magnetograms, taken in the Ni\,{\sc I} 676.77\,\rm{nm} line, have a cadence of 1\,min 
and a pixel size of 1.98\,$\arcsec$. 
The \emph{SOLIS-VSM} data is aligned to the \emph{MDI} magnetograms by cross-correlating the images.
\emph{RHESSI} recorded from 19:54 UT and the light curve is shown in Fig.~\ref{fig:rh_lc}. We reconstruct a low-energy X-ray
 image in the range 12\,keV to 18\,keV in the time from 19:57:00 to 19:57:18 UT,
 choosing detectors 3 to 8 and a pixel size of 1\,$\arcsec$. A second image is obtained 
between 20:04:00 and 20:04:18 in the energy range of 25\,keV to 50\,keV with detectors 
2 to 9. Both times the CLEAN-algorithm~\citep{2002SoPh..210...61H} is chosen using a maximum of 150 iterations. The alignment between \emph{RHESSI} and \emph{MDI} is reliable \citep{2007ApJ...656.1187F} apart from an unknown, but small, uncertainty in the \emph{MDI} roll angle. The timing of the \emph{MDI} and \emph{TRACE} images was retrieved from the level one data fits header. The \emph{TRACE} whitelight images have a pixel size of 0.5\,$\arcsec$  and were obtained for 19:36 UT and 20:10 UT. We used a nearest-neighbor algorithm to find \emph{MDI} intensity images closest in time to which the \emph{TRACE} images were then spatially aligned to. The \emph{MDI} images were shrunk in this process to take the different plate scales into account. For the \emph{SOLIS-VSM} images constructed from the Stokes profiles we used the observation start time of the first slit in the active region scan. To determine the timing of individual profiles in the \emph{SOLIS-VSM} scans, we add the product of the slit number times the observation time per slit to this start time.  
The \emph{RHESSI} instrument is described in detail in~\cite{2002SoPh..210....3L}, the \emph{MDI} in~\cite{1995SoPh..162..129S} and \emph{TRACE} in~\cite{1999SoPh..187..229H}.

%============================================================================================================================================
   %RESULTS

%============================================================================================================================================

\section{Results}\label{sec:results}

\subsection{Magnetic vector field changes at the delta spot} \label{sec:fullfov}
Preliminary inversions of this data set were presented by~\cite{2009ASPC..405..311F}. As the deteriorating polarization modulator affected a wide region in the eastern part of the field of view, it has not been possible to deduce global magnetic field vector changes for the active region as a whole. We focus in our analysis on flare III, an X1.4 flare, with a flare site located in the umbra. The observations of this flare site were fortunately less affected by the instrumental difficulties. 
Figure~\ref{fig:invpanels} shows a close-up view of the $\delta$-spot umbra recorded at times before and after flare\,{\sc III}. Solar north is in the direction of the positive y-axis and solar east is in the direction of the negative x-axis. The \emph{TRACE} whitelight image shows a light bridge originating just to the south of the negative polarity umbra and protruding into the umbra. This structure stays stable throughout the entire time sequence. The areas A and B  are locations of low continuum intensities indicating strong fields. They are also marked in the magnetogram and exhibit in addition low polarization signals during the entire time sequence. The profiles at locations P\,{\sc II} and P\,{\sc IV} (marked in the lower panel magnetogram in Fig.~\ref{fig:invpanels}) exhibit multilobed asymmetric Stokes $V$ profiles and are, in addition to the reference profile at P\,{\sc I}, examined in detail in section~\ref{subs:tempdev}. P\,{\sc II} is located at the polarity inversion line (PIL) showing multilobed asymmetric profiles throughout the entire time sequence whereas the profiles at P\,{\sc IV}, located in area A, only show asymmetric Stokes $V$ profiles during the rising phase of the flare. The magnetograms were obtained from Stokes $V$. Instead of computing Stokes $V$ at a wavelength position in the wing of the spectral line, we integrate the Stokes $V$ signal over $\lambda$. An antisymmetric spectral mask is first applied to the Stokes $V$ signal before the integration. Stokes $V$ signals with the same antisymmetry properties as the mask will therefore appear positive whereas the integration of the opposite Stokes $V$ signal will be negative. The units correspond then to the total area of the absolute Stokes $V$ signal.
The third and the fourth panel in Fig.~\ref{fig:invpanels} show the inferred magnetic field strength and the inclination angle $\gamma$ at optical depth log$(\tau_{\rm 500nm})={\rm -1}$ as retrieved with the inversion code LILIA.  The upper left areas in the inversion results are strongly affected by the noisy data and a `pixelation effect'  is clearly visible. 

To visualize any changes in the magnetic field vector we subtract the longitudinal magnetic field $B_{\rm long}$, the transversal field $B_{\rm trans}$ and the disambiguated azimuth angle of the magnetic field vector on the solar surface before and after flare\,{\sc III} (19:50 UT and 20:57 UT) and display the changes in Fig.~\ref{fig:azy}. 
On the left side upper panel in Fig.~\ref{fig:azy} one can see that the longitudinal magnetic field changes its sign mainly at the PIL separating the two umbrae (yellow contour on TRACE whitelight image). We use a cross correlation algorithm to align the scans taken at different times, after which only a slight jitter remains visible 
to the naked eye. Although a real horizontal motion of solar features would affect most prominently locations with strong magnetic field gradients, the fact that there are several regions located at the PIL in the penumbra surrounding the negative polarity umbra showing the same polarity switch identifies an instrumental effect e.g. a remaining pointing error and/or a change in seeing quality the most likely cause. The changes in azimuth angle (purple in the same image) are more diffuse. There is a larger connected patch located in the positive polarity umbra. The azimuth angle is measured counterclockwise from the solar north and as the azimuth increases in this area, it therefore indicates a counterclockwise movement of the solar surface projection of the magnetic field vector. In the third upper panel we have selected three regions R1, R2 and R3 at the PIL where we find changes greater than 400\,G in both transversal and longitudinal magnetic field. As the increase in one parameter is accompanied by a decrease in the other, these are areas of real directional change of the magnetic field vector.  As one can see in the second and third panels of the upper row in Fig.~\ref{fig:azy}, the previously mentioned area of azimuthal angle change in the umbra is co-spatial with a region in which the magnetic field vector becomes more horizontal and is part of R3.  
We plotted the average inclination angle $\gamma$ and the average $B_{\rm trans}$ in the regions R1, R2 and R3 in the last three rows in Fig.~\ref{fig:azy}. Location R1, at the tip of the lightbridge in the negative umbra, shows a decrease in $\gamma$ and, as $\gamma$ is above 90\,$\degr$, this constitutes a decrease in longitudinal magnetic field. This is accompanied by an increase in transversal magnetic field. The trend is visible in the entire time series. However, there seems to be a distinct outlier at 19:50 UT in region R1.  This is during the rising time of the soft X-ray flux (purple line). At that moment the field appears to become abruptly more vertical as can be seen also in the sudden decrease of transversal magnetic field. Region R2 is a patch across the entire PIL separating the umbrae. Here $\gamma$ and the longitudinal magnetic field increase, accompanied by a decrease in transversal magnetic field. In region R3 there is an overall trend in the time series showing an increase in $\gamma$, which corresponds to a decrease in $B_{\rm long}$ as ${\rm \gamma < 90\,\degr }$, and an increase in transversal magnetic field. There is a large variation between time steps with a zigzag pattern, leaving the impression of a swaying magnetic field vector. Region R3 is, however, located in the positive polarity umbra and is in the region affected by the instrumental difficulties mentioned in earlier sections. These profiles were more noisy, difficult to invert and the results have to be taken with caution.

\subsection{ \emph{SOLIS-VSM} Stokes profiles}
\label{subs:tempdev}
In Fig.~\ref{fig:allprofs} we plot examples of different types of Stokes profiles we encounter in the sunspot. Their locations (P\,I to P\,{\sc IV}) are indicated in the magnetogram in Fig.~\ref{fig:invpanels}.  In addition a quiet Sun profile (location not specified in Fig.~\ref{fig:invpanels}) is shown for comparison.

\subsubsection{General profile in the sunspot at P\,I:}
\label{gen}

The Stokes profiles located at P\,I, at the umbral edge, show broadened Stokes $I$ profiles. The FWHM is almost two times larger than in the quiet Sun (see Fig.~\ref{fig:allprofs}) obviously due to the Zeeman splitting in the strong umbral field. There is no significant $Q$ or $U$ signal, and one observes an \emph{anti}-symmetric Stokes $V$ profile as would emerge from a one-component magnetized atmosphere with the magnetic field vector along the line of sight and no large gradients in the magnetic and velocity field. We are able to easily fit such profiles with the inversion code, and the synthetic profiles are over-plotted with red lines.

\subsubsection{Asymmetric Stokes $V$ profile at polarity inversion line at P\,{\sc II}:}
\label{PILprof}
The Stokes profiles for both Fe\,{\sc I} lines located at P\,{\sc II} at the PIL, display a shallow intensity profile with a strong line broadening in Stokes $I$, a linear polarization signal in Stokes $Q$ and a weak, very asymmetric but not \emph{anti}-symmetric circular polarization in Stokes $V$. We find several such Stokes $V$ profiles located at the PIL. They maintain their asymmetric Stokes $V$ profiles throughout the entire time series. They can be readily explained as so-called crossover points.~\cite{1991sopo.work..307S} observed that asymmetric Stokes $V$ profiles found at the polarity inversion lines of $\delta$-spots are reproducible by combining `normal' \emph{anti}-symmetric opposite polarity Stokes $V$ profiles with small velocity differences ($\Delta \lambda$ in wavelength) from either side of the PIL. 
 A combination of observed profiles taken from either side of the PIL is qualitatively capable of reproducing the shape of the asymmetric profile at P\,{\sc II}. In addition the Stokes $Q$ and $U$ profiles stay the same in magnitude and shape  on either side of the PIL.

\subsubsection{Profiles in patch A and B in the negative and positive polarity umbra (P\,{\sc III} and P\,{\sc IV}):}
\label{lowpol}

The profiles at P\,{\sc III}, and P\,{\sc IV}, representing the pixels located in patch B and A, respectively, show a low continuum intensity and broadened Stokes $I$ profiles, as expected in areas of strong magnetic field. However, there is no linear polarization visible, and the Stokes $V$ profile is very weak (less than 30\,{\%} than in its surrounding). 
Possible explanations for such a low polarization signal are strong magnetic fields canceling each other within the resolution element or/and a high (possibly polarized) stray-light factor or/and local heating causing a change in the line profile. The continuum intensity is at 20\,$\%$ of the quiet Sun intensity at these locations. Previous studies of the umbral atmosphere at very strong magnetic field locations would suggest an intensity of $\sim$\,10\,$\%$ in our wavelength range~\citep{1970SoPh...13..312M}. On close inspection one finds that the splitting is more discernible in the Stokes $I$ profile of the {Fe\,{\sc I} 630.25\,\rm{nm}} line at P\,{\sc I} than in P\,{\sc IV} or P\,{\sc III} and one can differentiate between three distinct components in the Stokes $I$ profile. These two observations hint at a high stray-light factor in this region. 

The low polarization signature remains the same for the entire observing time in patch B. In patch A however, the polarization signal becomes even more reduced and we observe multilobed profiles in one of the {Fe\,{\sc I}} lines during flaring activity. The temporal evolution of the Stokes $V$ profiles located at  P\,{\sc IV} (patch A) is shown in Fig.~\ref{fig:temp1}(a). Between 19:47 and 20:07 UT, during the rising phase in the SXR flux of flare\,{\sc III} one can observe asymmetric multilobed profiles only in the {Fe\,{\sc I} 630.15\,\rm{nm}} line at PIV. The fact that only the Fe\,{\sc I} 630.15\,\rm{nm} line is affected is due to the line formation region difference between the two lines that is estimated (for non-magnetized regions) at 64\,km \citep{2009A&A...507L..29F} with the {Fe\,{\sc I} 630.15\,\rm{nm}} being formed in higher layers in the atmosphere. This means that the atmospheric parameters responsible for the observed asymmetry change their properties on a very short height range along the line of sight in the line formation region. If the low polarization signal and the shortlived multilobe profiles would be due to a real opposite polarity emerging we would expect the line with a formation region lower in the atmosphere ({Fe\,{\sc I} 630.25\,\rm{nm}}) to be more affected and as this is not the case we exclude this possibility.

We perform SIR inversions on the multilobed profiles taking only the spectral window of the {Fe\,{\sc I} 630.15\,\rm{nm}} line into account. We allow for two magnetized atmospheres within the resolution element and an unpolarized stray-light component. The best-fit synthesized profile for the Stokes $V$ and $I$ profile at location P\,{\sc IV} are shown in Fig.~\ref{fig:temp1}(b) and the corresponding retrieved atmospheric parameters are listed in Tab.~\ref{sirtab}. The Stokes $Q$ and Stokes $U$ signals were below the noise. The inversion results in a stray-light factor of 14.2$\%$. The two magnetized atmospheres exhibit a line-of-sight velocity difference of $\sim$\,12\,$\ensuremath{\mathrm{km}/\mathrm{s}}$ with an upflow of 4.64\,$\ensuremath{\mathrm{km}/\mathrm{s}}$ and a downflow of 7.35\,$\ensuremath{\mathrm{km}/\mathrm{s}}$. We were not able to invert some of the pixels in patch A and for this cases we used a simple fitting approach employing constructed gaussians as described in section~\ref{asymfitSIR_GAU}.

%======================================SIR INVERSION: TABLE
\begin{table}
\caption{Atmospheric parameters for the two magnetic atmospheres from the SIR inversion: line-of-sight velocity $v$, magnetic field $B$, temperature $T$, inclination angle $\gamma$ and filling factor $f$. The Stokes $V$ fit is shown in Fig.~\ref{fig:temp1} (b) with the Stokes $V$ profile in green being the profile synthesized from umbra model 1 and the Stokes $V$ profile in blue emerging from umbra model 2.
Negative line-of-sight velocity values correspond to an upflow.}
\label{sirtab}
\centering

	 	\begin{tabular}{l c c c c c }
Model&v [km\,s$^{-1}$]&B [G]&T [K]&$\gamma$ [$\degr$]&f\\
  \hline
 \hline
 Umbra 1&7.35&1050&4046&166&0.2\\
 Umbra 2&-4.64&1392&3124&178&0.8\\

 \hline	

\end{tabular}
\end{table}
%=====================================

In Fig.~\ref{fig:temp1}(c) we show such a fit for a pixel also located in Patch A and close to P\,{\sc IV}. Depending on the start model (same polarity or opposite polarity atmospheres) we arrive at different fit solutions and obtain two possible atmospheric scenarios that could cause the asymmetric Stokes $V$ profile. The profile can be reproduced by two same polarity synthetic Stokes $V$ profiles with a high velocity difference. The best fit achieved in this way produces a line-of-sight velocity difference of $\sim$\,13.3\,$\ensuremath{\mathrm{km}/\mathrm{s}}$. Calibrating the wavelength scale with the telluric line position, the blue-shifted line has a velocity of  $\sim$\,8.2\,$\ensuremath{\mathrm{km}/\mathrm{s}}$ and the red-shifted exhibits a velocity of $\sim$\,5.1\,$\ensuremath{\mathrm{km}/\mathrm{s}}$. The second possibility is a locally heated atmosphere causing the complex emerging Stokes $V$ profile as has been found previously by e.g.~\cite{2000Sci...288.1396S} in their umbral flash studies using the Ca\,{\sc II} infrared triplet. They observed shortlived asymmetric Stokes $V$ profiles in the sunspot umbra and were able to show that these are caused by the combination of two unresolved atmospheric components, one resulting in a normal Stokes $V$ profile and the second producing a blueshifted reversed Stokes $V$ signal due to an emission process taking place. In the second row in  Fig.~\ref{fig:temp1}(c) we demonstrate this possibility with our simple gaussian model.
We fit the profile using a Stokes $V$ signal as it would arise due to absorption and a reversed polarity signal caused by an emission feature. The reversed Stokes $V$ profile exhibits a narrower profile which could be due to the emission feature being present mainly in the core of the spectral line. The line-of-sight velocity difference between the two \emph{anti}-symmetric profiles is $\sim$\,0.48\,$\ensuremath{\mathrm{km}/\mathrm{s}}$. 

As the multilobed profiles appear in a strong magnetic field location with a high stray-light percentage in which the polarization signal is already reduced in amplitude, the effect of the spectral line going into emission is visible in Stokes $V$ whereas the line core reversal can be completely masked in Stokes $I$ as was demonstrated in~\cite{2011NV}. They synthesized profiles from MURaM snapshots \citep[see code description]{2003AN....324..399V} and showed that the expected line reversal in the Stokes $I$ profile emerging from an atmosphere with a temperature increase with height was masked by adding a small stray-light contamination whereas the Stokes $V$ profiles maintained a complex shape.

In both cases, large velocity difference and local heating scenario, a good fit is achieved using two \emph{anti}-symmetric Stokes $V$ profiles. This indicates a multicomponent magnetic atmosphere and sub-resolution fine structure.

%MDI ANALYSIS
\subsection{Sign-reversal in \emph{MDI} magnetograms and RHESSI reconstruction}
\label{subs:mdirev}

%======================================FIGURE 10: MDI PANELS
\begin{figure*}
\centering
  \includegraphics[width=17cm]{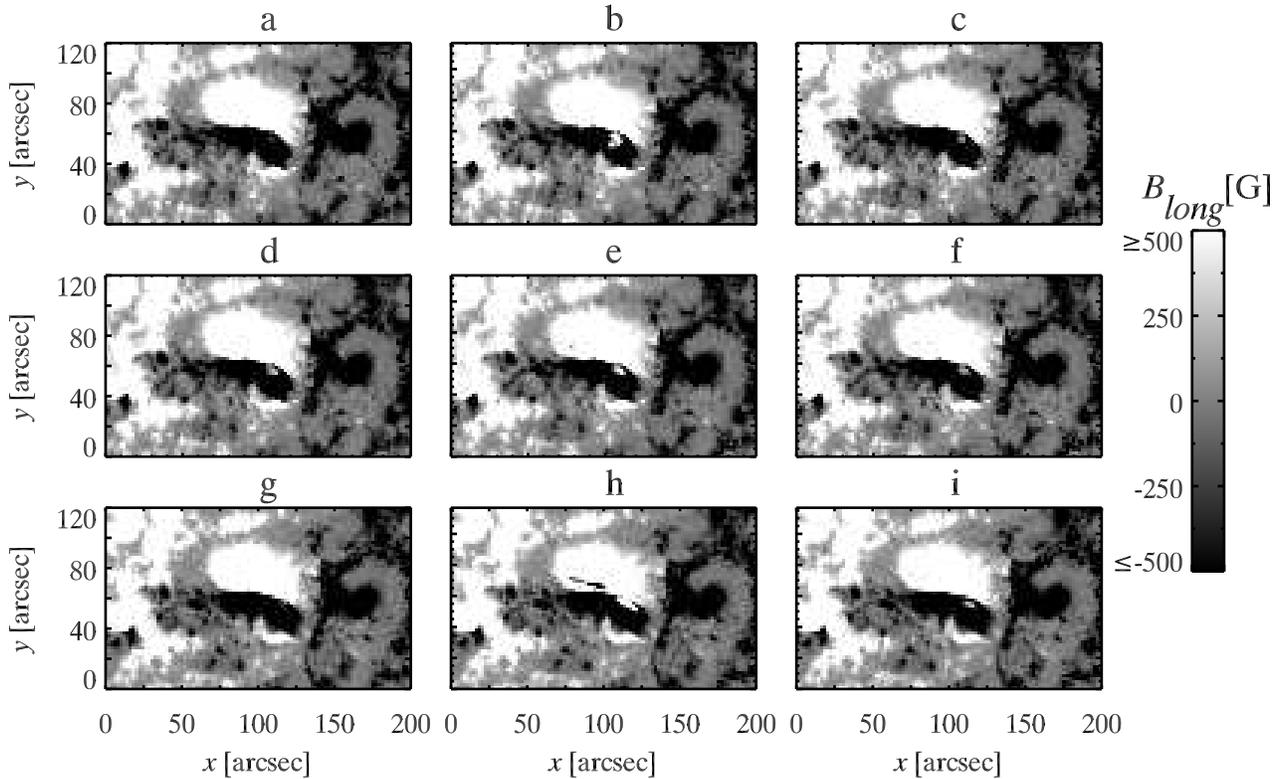} 
 \caption[]{
    \emph{MDI} magnetograms during the flares\,{\sc I},\,{\sc III} and\,{\sc V}. The images are all clipped to the same maximum and minimum of $\pm$500\,G. The observation times are indicated by the vertical lines in Fig.~\ref{fig:utp} corresponding to the letters a to i. {\em Upper row [a,b,c]:\/} The first panel shows the sunspot before flare\,{\sc I}. Panel two was taken during the rising phase of flare\,{\sc I} and panel c after the flare SXR peak. {\em Middle row [d,e,f]:\/} Magnetograms before flare\,{\sc III}, during the rising phase of the flare and after.  {\em Lower row: [g,h,i]\/} Magnetograms showing the sunspot this time before flare\,{\sc V}, during the rising phase of the flare and again after.}
\label{fig:mdialone} 
 \end{figure*}
%===================================================================================
%======================================FIGURE 11 : UT-PLOT OF WHITE PATCH
\begin{figure*}
\centering
\includegraphics[width=17cm]{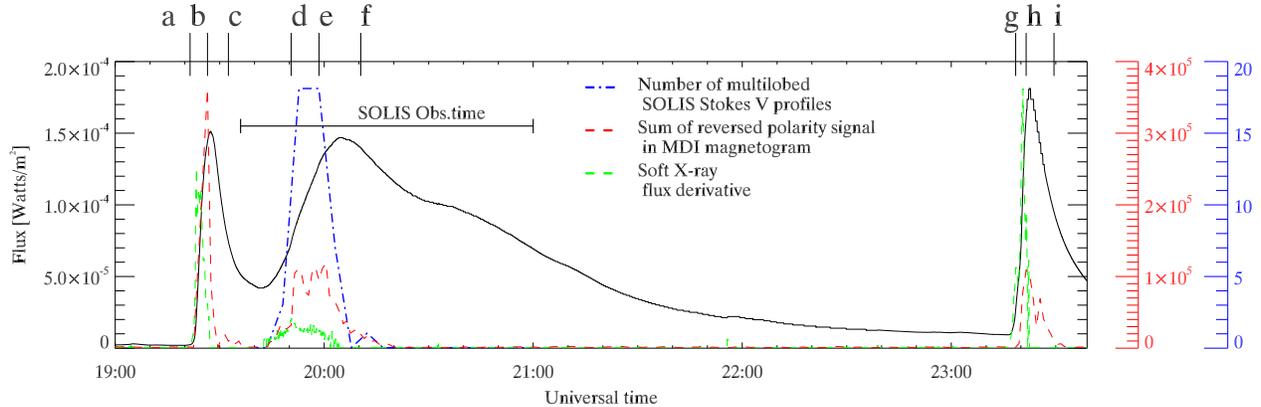} 
 \caption[]{Soft X-ray flux recorded by \emph{GOES} between 19:00 UT and 23:40 UT (solid line). The dashed green line corresponds to its derivative. The dashed red line corresponds to the sum of parasitic positive magnetogram values found in patch A in the negative polarity umbra of the \emph{MDI} magnetograms. The function was multiplied by a factor for the purpose of shifting the values into the plotting range. The blue dashed-dotted line is the number of multilobed Stokes $V$ profiles observed with \emph{SOLIS} and found in patch A again scaled to the plotting window. The vertical lines at the top of the Figure signify the record time of the magnetograms (a-i) displayed in Fig.~\ref{fig:mdialone}.  } 
 \label{fig:utp}
 \end{figure*}
%===================================================================================

%=================================================FIGURE 12 RHESSI PANELS
\begin{figure*}
\sidecaption
  \includegraphics[width=12cm]{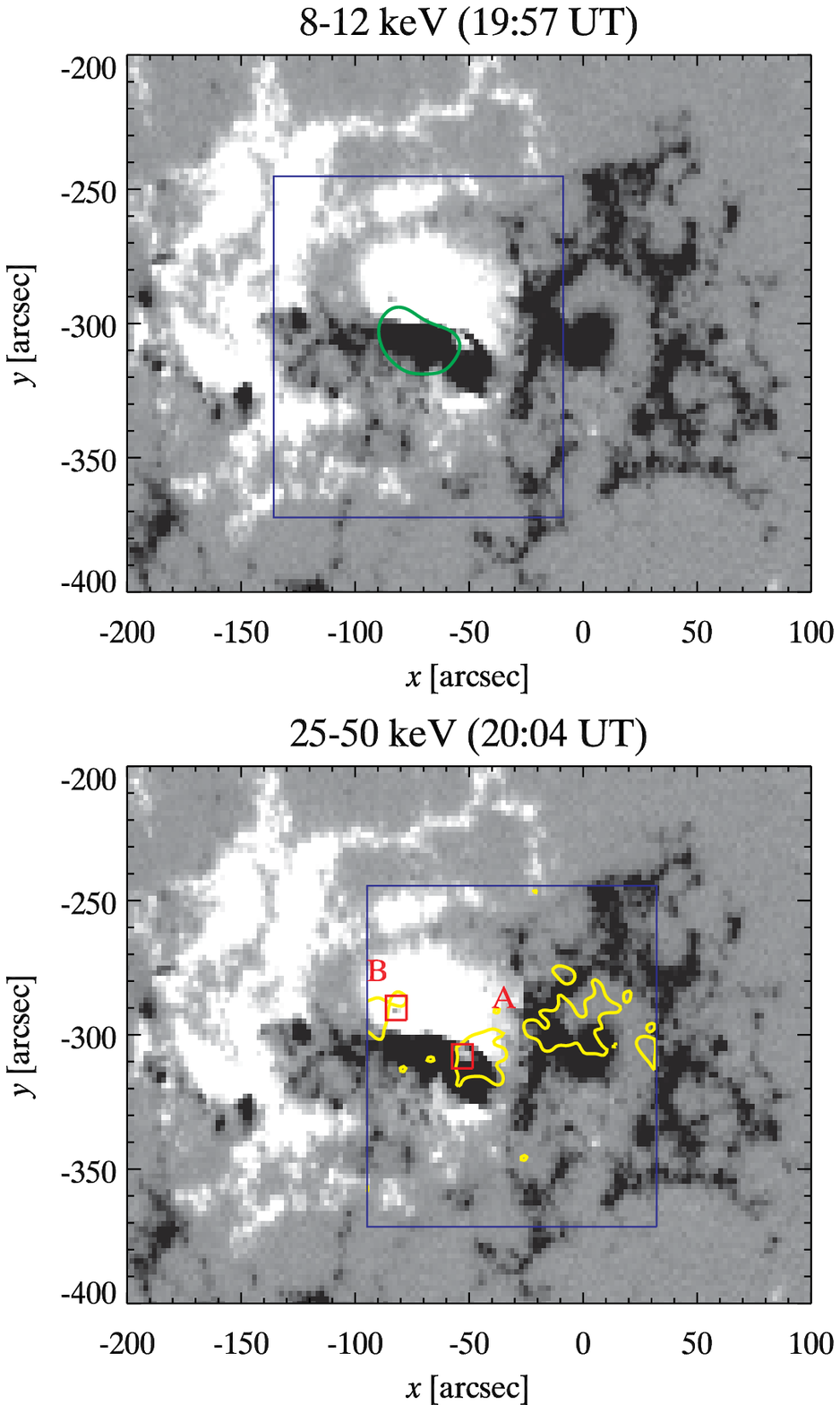} 
  \caption[]{
 \emph{RHESSI} reconstructions during flare\,{\sc III}. The times refer to the MDI magnetogram. For the RHESSI reconstruction times refer to section~\ref{cdata}.
  {\em Upper row:\/}  \emph{RHESSI} low-energy X-ray reconstruction (12-18\,\ensuremath{keV}) contoured at a level 
of 40 \% and overlaid onto a \emph{MDI} magnetogram with the same cutoff level of $\pm$500\,G as in Fig.~\ref{fig:mdialone}. The low-energy X-ray emitting region is located above the PIL at the $\delta$-spot. The blue box is the FOV chosen for the \emph{RHESSI} reconstruction.
{\em Lower row:\/} \emph{RHESSI} hard X-ray reconstruction (25-50\,\ensuremath{keV}) contoured at a level
 of 40 \% and overlaid onto a \emph{MDI} magnetogram. The blue box is again the FOV chosen for the \emph{RHESSI} reconstruction. The regions denoted A and B are the same as in Fig.~\ref{fig:invpanels}.}
 \label{fig:trar} 
\end{figure*}
%===========================================================================
We select \emph{MDI} magnetograms before, during and after the \emph{GOES} SXR flux peaks of flares\,{\sc I},\,{\sc III} and\,{\sc V}. The magnetograms are shown in Fig.~\ref{fig:mdialone} and are labeled {\em a} to {\em i}. The times corresponding to these letters can be found in Fig.~\ref{fig:utp}. During the soft X-ray flux rising times of flare\,{\sc I} (panel {\em b} in Fig.~\ref{fig:mdialone}) one can observe a patch (co-spatial with patch A from Fig.~\ref{fig:invpanels}) exhibiting an apparent reversed polarity in the negative umbra, before being reduced in amplitude shortly afterward (panel {\em c}). 
During flare\,{\sc III} the reversed polarity, located at patch A, is more prominent during the SXR flux rising phase (panel e). The last flare shows an elongated structure exhibiting strong sign reversal and spanning over the PIL connecting patches A  and B.

\emph{MDI} uses the difference of dopplergrams observed in the Stokes $I-V$ and $I+V$ profiles of the Ni\,{\sc I} 676.77\,\rm{nm} line to quantify the magnetic field instead of Stokes $V$ amplitude maxima as measured by a classical magnetograph ~\citep{1995SoPh..162..129S}.  \emph{MDI}-saturation, showing up as grey patches in the magnetogram, is  caused by the low intensities measured in the strong umbra rendering the on-board processing algorithm ineffective as shown by \cite{2007SoPh..241..185L}. A polarity reversal, observed as a parasitic opposite polarity in an otherwise unipolar field however hints at a real physical meaning. If actual flux emergence is excluded, other possibilities are supersonic speeds, pushing the spectral line partly out of the analysis window or line reversal due to the line going into emission~\citep{2003ApJ...599..615Q}.

To investigate the timing relationship between the reversed polarity, multilobed Stokes $V$ profiles observed with \emph{SOLIS} and the flare excitation we plot the total intensity in the pixels exhibiting the reversed polarity and the number of multilobed Stokes $V$ profiles as a function of time, and compare with the \emph{GOES} SXR flux, and its time derivative. The latter can usually be taken as a good proxy for the HXR time profile~\citep[via the 'Neupert Effect', see e.g.,][]{1993SoPh..146..177D} which was not continuously available throughout the event. As shown in Fig.~\ref{fig:utp} the timing of this reversed polarity and the unusual Stokes $V$ profiles coincides very well with the derivative of the soft X-ray flux. The appearance of the apparent reversed polarity and the multilobed Stokes $V$ profiles is therefore very likely due to impulsive-phase, non-thermal processes which must, therefore, be affecting the photospheric layers. 

This is supported by the \emph{RHESSI} reconstructed images shown in Fig.~\ref{fig:trar}. The data was taken during the rising SXR flux phase (see section~\ref{sec:dataana} for processing details). In our reconstructions~\citep[in accordance with the findings of][]{2009ApJ...703..757L} we can identify a low-energy X-ray source over the negative polarity umbra and stretching on one side over the PIL into the positive umbra and at least two non-thermal HXR sources placed at the location of patches A and B. In a classic flare picture the SXR is a loop top source of flare {\sc III}  with the hard X-ray sources located at the chromospheric footpoints of the interacting loops. These footpoints are areas with high-energy electrons impacting the lower atmosphere.

%============================================================================================================================================
   %DISCUSSION/CONCLUSION
   
%============================================================================================================================================

\section{Discussion and Conclusion} \label{sec:discussion}

We applied a height-dependent inversion to the full Stokes profiles data of flare\,{\sc III} and find a change of magnetic field vector direction along the polarity inversion line. We identify regions R1 and R3 which show an increase in transversal magnetic field accompanied by a decrease in longitudinal magnetic field and a region R2 in between showing the opposite movement (see Fig.~\ref{fig:azy}). These regions are located close to the PIL separating the umbrae. With their extensive multiwavelength analysis \cite{2009ApJ...703..757L} have speculated on the flaring loop configuration (see their Fig. 7 panel d) of flare\,{\sc III}. Two of the flare kernels (their k1 and k2, corresponding to our patch A and B) are in the positive and in the negative polarity umbra respectively and are connected to kernels outside the $\delta$-spot. According to their speculation the flare is a consequence of reconnection taking place between these loop systems just above the PIL separating the two umbrae. After the flare k1 and k2 are connected by a low lying loop. This scenario would fit well with our observations, as the transverse magnetic field increases on the opposite sides of the PIL (regions R1 and R3 in Fig.~\ref{fig:azy}) indicating the development of a lower lying loop. The increase in longitudinal magnetic field at location R2 could be attributed to the removal of opposing longitudinal fields close to the X-point above the PIL. 
This is a favorable two-loop configuration in the flare model of~\cite{1997ApJ...486..521M} where a longer loop and a low-lying shorter loop are formed as a consequence of reconnection between two like-polarity, collinear, current-carrying loops crossing each other above the photosphere.
Permanent changes of the magnetic field have been quantified in many flares. \cite{2005ApJ...635..647S} used \emph{GONG} magnetograms to identify magnetic field changes and found that the changes were often abrupt (less than 10 minutes) and permanent.  Also more recent observations  from~\cite{2012ApJ...745L..17W} taken with the \emph{SDO/HMI} instrument show an enhancement in horizontal field by 30\,$\%$ taking place within $\sim$\,30 minutes in an area close to the PIL during an X\,2.2 flare. Our findings show that the changes occur during almost the entire time sequence (90 minutes) and are therefore not sudden. In addition flare\,{\sc III} occurs after two previous flares, and we do not have enough coverage before and after the flares to compare the field changes in the same area when not affected by a flare.

We find two locations of strong magnetic field A and B, as inferred from the corresponding Stokes $I$ profiles, that show a low polarization signal (less than 30\,{\%} than in their surrounding). Profiles located in patch A, such as P\,{\sc IV}, also show multilobed asymmetric Stokes $V$ profiles during flare activity co-spatial and co-temporal with a reversed polarity signal in \emph{MDI} magnetograms and a hard X-ray source observed with \emph{RHESSI}. The asymmetry seen in Stokes $V$ lasts around 15 to 20 minutes and occurs during the impulsive phase of flare\,{\sc III} and only in the {Fe\,{\sc I} 630.15\,\rm{nm}} line. As the {Fe\,{\sc I} 630.25\,\rm{nm}} line is not affected and is formed around 64\,${\rm km}$ deeper than the {Fe\,{\sc I} 630.15\,\rm{nm}} line, this indicates a very rapid variation of the atmospheric properties with depth. 

The temporary reversed magnetic polarity patch, if indeed due to an actual magnetic feature rather than the disturbance of the spectral lines caused by flare heating, is reminiscent of the magnetic configuration described by~\cite{2000ApJ...528.1004F}. In this, an isolated patch of one 'included' polarity surrounded by the opposite polarity gives rise to a coronal null and a  'spine-fan' magnetic topology - which is also central to flares studied by~\cite{2000ApJ...540.1126A} and~\cite{2001ApJ...554..451F}. However, unlike in these models the reversed polarity patch in our flare is not stable, and does not persist after the flare, but is a transient feature. Also, due to the fact that the multilobe profile is observed only in the Fe\,{\sc I} 630.15\,{\rm nm} line it is also reasonable to exclude flux emerging at these times.

Resulting from our fits we find at least two possible atmospheric scenarios that are capable of reproducing these profiles. One is a two component atmosphere exhibiting a velocity difference between the components in the range of 12\,$\ensuremath{\mathrm{km}/\mathrm{s}}$ to 14\,$\ensuremath{\mathrm{km}/\mathrm{s}}$ or alternatively there is a polarity reversed contribution to the emerging Stokes $V$ profile due to a temperature increase from the local back-warming from the chromosphere. As this effect can be masked in Stokes $I$ due to a small filling factor of the reversed signal and a high stray-light factor at this location we were not able to distinguish between these two cases.

Observation of high velocities deduced from asymmetric Stokes $V$ profiles in a  sunspot were also found by~\cite{1994ApJ...425L.113M} and~\cite{2003A&A...412..541M}. In both cases they were observed close to the PIL in a $\delta$-sunspot.~\cite{1994ApJ...425L.113M}  Stokes $V$ fit showed velocities of 14\,$\ensuremath{\mathrm{km}/\mathrm{s}}$. However these were long-lasting flows and in the observations of~\cite{1994ApJ...425L.113M}  seemingly unconnected to a flare. 
A transient flow of 2\,$\ensuremath{\mathrm{km}/\mathrm{s}}$ at flare footpoints was reported recently by \cite{2011SoPh..269..269M} deduced from line shifts in the intensity profiles observed with the \emph{HMI} onboard \emph{SDO}  during a whitelight flare. A downflow of 8\,$\ensuremath{\mathrm{km}/\mathrm{s}}$ was found in the Na\,D$_{1}$ line (lower chromosphere) by \cite{2005ApJ...630.1168D}, in association with the launch of a flare seismic wave, and transient downflows have also been reported earlier by \cite{2006SoPh..238....1K}. Indeed, a popular model for launching the flare seismic waves is strong local heating in the chromosphere leading to downward-propagating hydrodynamic shocks  as found by  \cite{1985ApJ...289..434F}, which deliver energy and momentum to the photosphere. 

Apparent flux reversals in magnetograms during flares have been reported earlier by, e.g.,~\cite{1984ApJ...280..884P} and \cite{2003ApJ...599..615Q}. \cite{2003ApJ...599..615Q} also found co-spatial HXR sources at the reversed polarity sites. They simulated the \emph{MDI} signal in Ni\,{\sc I} 676.77\,\rm{nm} and suggest that the signal is caused by the line center or even the whole line going into emission. They find an interesting connection between locations of strong magnetic field and reversed polarity sites and point out that this could be due to the fact that only at these locations is the energy requirement to produce the effect low enough. Our case supports this as we find the asymmetric Stokes profiles only at the site of lowest continuum intensity in the umbra (see Fig.~\ref{fig:invpanels} green contours on \emph{TRACE} whitelight image) which therefore favors the local heating scenario. ~\cite{2012ApJ...747..134M} have observed an apparent reversed polarity in \emph{HMI} magnetograms during a 2.2 X-flare. They obtained the spectral line profile (6 spectral points) and find an intensity enhancement near the line core correlated temporally to the impulsive phase of the flare. Similar observational signatures are seen therefore in several spectral lines from {Ni\,{\sc I} 676.77\,\rm{nm}}(\emph{MDI}) to {Fe\,{\sc I} 617.33\,\rm{nm}} (\emph{HMI}) and {Fe\,{\sc I} 630.15\,\rm{nm}} (\emph{SOLIS-VSM}).

Our analysis confirms the direct connection between the chromospheric flare response and the photospheric atmosphere as seen in the Stokes profiles. Although we cannot differentiate between the two physical explanations, it is clear that the flare impact sites show a fine structure in the photosphere. Although a fine structure in the umbra, as we observed, seems surprising, this has been already suggested by e.g.,~\cite{2000ApJ...544.1141S} who studied asymmetric Stokes $V$ profiles in the infrared Ca\,II\ lines. They observed very similar short lived unusual Stokes $V$ profiles and were only able to invert the profiles with a sub-arcsec multicomponent atmosphere in the sunspot umbra. With the sub-arcsec resolution of instruments such as \emph{HINODE} it has been possible to actually observe this fine structure directly in the chromosphere~\citep{2009ApJ...696.1683S}.

Full Stokes profile observations during a flare with high spatial, spectral and temporal resolution probing, in addition to the photosphere, also the higher atmospheric layers through multi-wavelength observations are needed to be able to differentiate between the possible atmospheric scenarios leading to the anomalies seen in longitudinal magnetograms and asymmetric Stokes $V$ profiles. However, due to the current spatial resolution and the unpredictability of flare onset, the study of flare triggering mechanism, flare impact and photospheric response is still challenging. We have shown that fine structure is also important in areas of strong magnetic fields as encountered in sunspot umbrae and that the high resolution study of flare footpoints offers the possibility of understanding the coupling between the atmospheric layers in flare processes.

%============================================================================================================================================
   %THANKS

%============================================================================================================================================
\subsection{Acknowledgements}
We thank Aimee Norton for her correspondence concerning \emph{MDI} saturation features. 

This research project has been supported by a Marie Curie Early Stage
Research Training Fellowship of the European Community's Sixth Framework
Program under contract number MEST-CT-2005-020395.

L. Fletcher acknowledges support from STFC Grant ST/1001808 and the EC-funded FP7 project HESPE (FP7-2010-SPACE-1-263086).

The project has been also supported by the SOLAIRE network of the European
Community's Sixth Framework Programme.

%============================================================================================================================================
   %REFRENCES

%============================================================================================================================================

\parskip=-1ex
\bibliographystyle{aa}   
\bibliography{cef19272}

\end{document}